# Synthesis and characterization of new chitosan-based nanocomposite gel and its application towards dye removal

*Tamal Sarkar**

Raman Research Institute, Bangalore, India

*Email: tamalsarkar09@gmail.com

## Abstract

We present a nanocomposite gel that efficiently adsorbs a toxic and non-biodegradable cationic dye (methylene blue) from water. This nanocomposite gel is synthesized using chitosan, silica, and titanium dioxide nanoparticles ($TiO_2$ NPs) as precursors. The structure of the nanocomposite gel is characterized by X-ray diffraction (XRD) and cryogenic-scanning electron microscopy (cryo-SEM). Its viscoelastic properties are analyzed using a rheometer, and the mechanical rigidity is noted to significantly improve upon the incorporation of $TiO_2$ NPs. The nanocomposite gel is then utilized to adsorb methylene blue dye (1-20mg/L) from its aqueous solution using batch method. Effects of contact time, pH of the solvent, amount of adsorbent dose, and initial dye concentration on the dye removal percentage are probed systematically. The kinetic studies indicate that the adsorption process follows pseudo-second-order kinetics. Among various isotherms, Freundlich isotherm describes the adsorption best with the highest regression coefficient. Furthermore, preliminary results show that the nanocomposite synthesized here even has the capability of adsorbing heavy metal ions such as chromium (Cr(VI)). This nanocomposite gel is, therefore, a promising candidate for the treatment of wastewater containing dyes and other contaminants from wastewater.

## 1. Introduction

In this time of ever-increasing industrialization, one of the most significant concerns regarding our environment is the contamination of water by organic/inorganic species, heavy metal ions, etc. [1][2]. There are several techniques for the treatment of wastewater, such as adsorption, solvent extraction, precipitation, membrane filtration, and ion exchange. Adsorption is relatively advantageous because it needs minimum usage of chemicals, is low cost, and the process is flexible [2]. Biosorption, a subtype of adsorption, is an emerging technology where metal ions and organic/inorganic species are absorbed from solution and assimilated by biomaterials [3]. Biosorption has several advantages over conventional methods due to biodegradability, biocompatibility, and cost-effectiveness of the biomaterials, recovery of the biomaterials after use, a smaller requirement for chemicals, the abundance of biomaterials in nature, and possible reuse of the recovered elements [4][5][6].

Chitosan, a biopolymer available abundantly and usually derived from deacetylated chitin, is an excellent biosorbent. For biosorption, chitosan has some specific advantages such as reusability, high concentration of amino groups, ease of functionalization, etc. Chitosan, therefore, shows superior adsorption properties and selectivity towards metal ions and organic species [7][8][9]. However, chitosan does have certain limitations, such as poor mechanical properties, low chemical resistance, swelling, and solubility in acidic conditions. For its use in wastewater treatment, chitosan has been modified with several materials such as alginate, alumina, polyvinyl alcohol, cyclodextrins, magnetite nanoparticles, ionic liquid, and silica [10]. Such modification ensures the availability of several useful properties of both the nanomaterial and the hydrogel in their nanocomposites.

Owing to the large surface area, high porosity, stability over a wide pH range, and the excellent mechanical resistance of chitosan-silica gels [11], they are considered promising candidates for several applications such as high-performance liquid chromatography, adsorption, etc. Silicas are characterized by high surface area, thermal stability, resistance to microbial attack, usability as an organic solid-support, and low cost [12]. Hassan *et al.* reported the use of a chitosan-silica composite to immobilize zinc oxide (ZnO) nanoparticles into a chitosan/silica/ZnO nanocomposite [13].

Titanium dioxide nanoparticles ($TiO_2$ NPs) have far-reaching uses as an excellent photocatalyst due to their high photocatalytic activity, photostability, reasonable price, and chemical and biological activity [14]. As $TiO_2$ NPs show very little selectivity, they are able to degrade almost all kinds of contaminants such as dyes, heavy metals, chlorinated organic compounds, phenols, pesticides, arsenic, and cyanide [15]. One major disadvantage of $TiO_2$ NPs is that it is very difficult to separate or recover $TiO_2$ NPs from the treated water when used in solution. Several methods have been proposed to address this problem. Among them are coupling $TiO_2$ NPs with a membrane [16], nanocomposite hydrogels using polymers [17], etc. For example, methylene blue was removed from wastewater by $TiO_2$ NPs in the form of $TiO_2$/poly[acrylamide-co-(acrylic acid)] hydrogel [17]. The incorporation of nanomaterials into polymers or hybrid networks also improves the mechanical rigidity of the composite materials [18][19].The improved rigidity of the adsorbent is essential for wastewater treatment because removal of the adsorbed contaminants by the adsorbent is crucial after the treatment [20]. In the present study, the improvement of the rigidity of the nanocomposite after incorporation of $TiO_2$ NPs is verified using a systematic rheology study.

We report here a systematic study on a versatile and easily-prepared chitosan-based nanocomposite gel that also shows excellent promise in the removal of certain cationic contaminants from water. In this study, we synthesize a chitosan-silica matrix that homogeneously traps $TiO_2$ NPs using sol-gel polymerization method. The formation of the composite gel is characterized by X-ray diffraction and cryogenic scanning electron microscopy (cryo-SEM) studies. The improvement in the mechanical rigidity of the hybrid gel after incorporation of TiO$_2$ NPs is shown by the rheology study. The chitosan-silica-TiO$_2$ nanocomposite shows excellent removal capacity of methylene blue dye (MB) from water in the concentration range 1-20 mg/L. Given the low selectivity of TiO$_2$ in the adsorption of contaminants, the chitosan-silica-TiO$_2$ nanocomposite, therefore, shows great promise in the treatment of wastewater. Finally, we have also reported a very preliminary study to demonstrate the removal of Cr(VI) ions by this nanocomposite.

## 2. Experimental Methods

### 2.1 Materials

Tetraethyl orthosilicate (TEOS), chitosan (molecular weight 190,000-375,000Da), titanium dioxide (TiO$_2$) NPs, and methylene blue and potassium dichromate were procured from Sigma Aldrich, Germany. Ethanol was purchased from Merck, Germany. Hydrochloric acid (HCl) was bought from SDFCL, India. All the glassware was purchased from Borosil, India.

### 2.2 Synthesis of chitosan-silica hybrid and chitosan-silica-TiO2 nanocomposite gel

The chitosan-silica hybrid gel was synthesized by the sol-gel method [21], which typically consists of hydrolysis followed by condensation [18]. Tetraethyorthosilicate (TEOS) was used as the silica precursor and hydrolyzed in the presence of a chitosan solution. The technique is as follows: 0.5% chitosan (10ml) was added to a silica solution. The silica solution was prepared by vigorously mixing 4.6ml TEOS, 3ml ethanol, 0.5ml HCL and 1ml H$_2$O (pH~1). The resulting mixture was stirred for 1h and kept in an oven at 60$^o$C. After 1 day, the sol transformed to a chitosan-silica hydrogel.

To synthesize chitosan-silica-TiO2 nanocomposite gel, TiO$_2$ NPs were incorporated in the hybrid chitosan-silica system before the occurrence of gelation. TiO$_2$ NPs were added to the TEOS solution, and the resultant solution was then added to a chitosan solution. The chitosan-silica-TiO$_2$ sol was vigorously stirred for 1h to achieve homogeneity in the diffusion of TiO$_2$ NPs. The mixture was kept inside the furnace at 60$^o$ C. The synthesis of both the chitosan-silica and chitosan-silica-TiO₂ hybrids are presented in the schematic diagram [Fig.1].

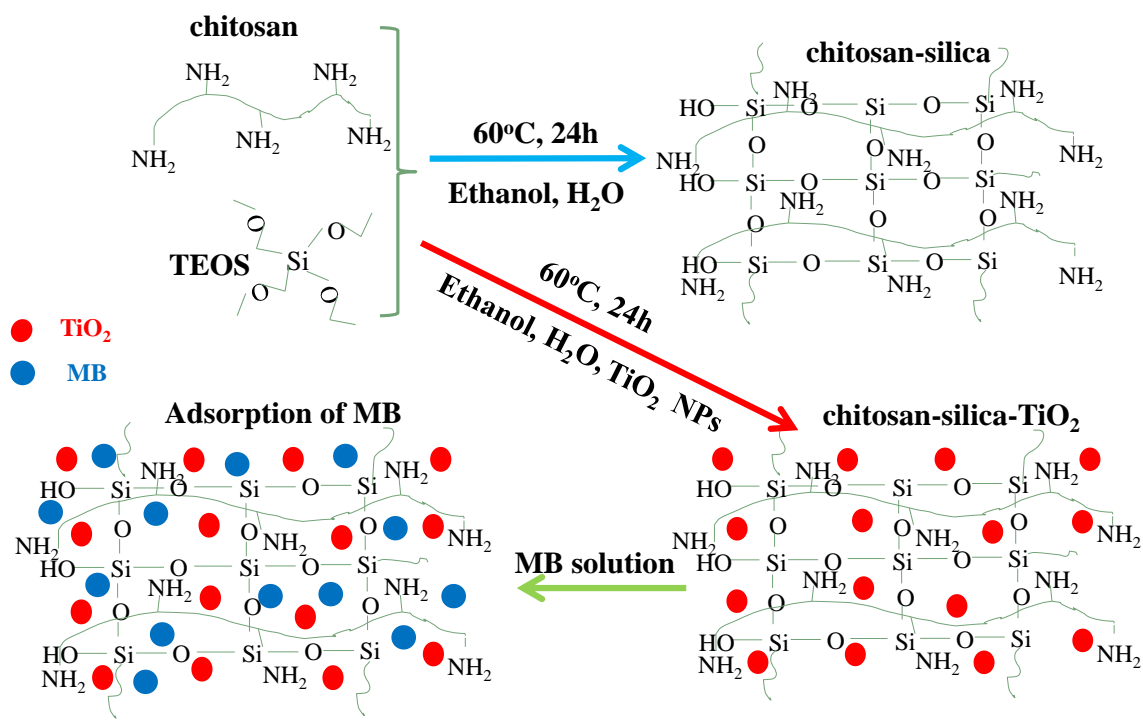

**Figure 1:** Schematic representation of the synthesis of chitosan-silica-TiO$_2$ nanocomposite gel and adsorption of methylene blue.

### 2.4 Characterization of the nanocomposite gel

The presence of TiO$_2$ NPs inside the chitosan-silica composite gel was confirmed by performing an X-ray diffraction study (PANalytica, UK). Network formation of silica inside the chitosan layers was observed in a cryo-SEM study (Carl Zeiss, Germany). Energy dispersive X-ray (EDX) spectroscopy was employed to confirm the presence of TiO$_2$ in the chitosan-silica matrix. All the adsorption studies were performed using a UV-Vis spectrometer (Perkin Elmer Lambda 35, USA). Rheological studies were performed to probe the viscoelastic properties of the nanocomposite gel using a stress-controlled rheometer (Physica MCR 501, Anton Paar, Germany) with a cone and plate (CP-25) measuring geometry of cone diameter 25 mm and cone angle 0.979°. In this work, two oscillatory procedures have been utilized, *viz.* strain amplitude sweep and frequency sweep measurements. In an amplitude sweep measurement, the strain amplitude $\gamma_0$ of an applied oscillatory strain $\gamma = \gamma_o \sin \omega t$ is varied at a constant frequency $\omega$ and the elastic and viscous moduli, $G'$ and $G''$, are measured. In a frequency sweep measurement, $G'$ and $G''$ are measured while the frequency $\omega$ is varied and the oscillatory strain amplitude $\gamma_o$ is kept constant.

### 2.5 Adsorption studies

For demonstrating the adsorption of dye by chitosan-silica-TiO$_2$ nanocomposite gel, batch method was used here. All adsorption studies were conducted at room temperature (25°C). The effects of pH, contact time, and nanocomposite dose concentration on adsorption were studied. Adsorption capacity $q_c$ (mg/g) of the chitosan-silica-TiO$_2$ nanocomposite hydrogel and removal percentage of the MB dye (in the concentration range 1-20mg/L) were determined using the following equations [13]

$q_c = (C_0 - C_e)V/m$ (Eq. 1)

Removal % $= ((C_0 - C_t)/C_0) \times 100$ (Eq. 2)

where, $C_0$ (mg/L) is the MB dye concentrations at the initial time, $C_e$ (mg/L) is the concentration at which removal rate slows down abruptly (identified as an approximate equilibrium value) and $C_t$ (mg/L) is the concentration at a contact time $t$. $V$ (L) is the total volume of the MB dye solution, and $m$ (g) is the mass of the chitosan-silica-TiO$_2$ nanocomposite gel. The determination of removal % was implemented using systematic steps. For a representative initial MB concentration of 5mg/L at a pH of 7, the protocol for the estimation of the removal % after adsorption by the chitosan-silica-nanocomposite at pH-7 is described in section S.A. in Supporting Information.

### 3. Results and Discussion

### 3.1. X-ray diffraction analysis

XRD patterns of chitosan, TiO₂ NPs, chitosan-silica hybrid, and chitosan-silica-TiO$_2$ nanocomposite gel are shown in Fig. 2. The chitosan powder shows distinguishable peaks at 2θ values of 9.39º and 19.63º (Fig. 2(a)). TiO₂ NPs exhibit crystalline peaks at 2θ values of 25.38º, 38.02º, 48.26º, 54.12º, 55.32º, 62.85º, 69.01º, 70.52º, 75.19º, 83.01ºthat are assigned to the crystalline planes (101), (004), (200), (105), (211), (204), (116), (220), (215), and (224) respectively [22], presented in Fig. 2(b).Fig. 2(c) shows the XRD pattern of the chitosan-silica hybrid, which exhibits no distinct peak besides a broad one at 23º, which is attributed to the presence of polysaccharide diffraction [13]. The chitosan-silica-TiO₂ nanocomposite (XRD pattern in Fig. 2(d)) exhibits all the peaks ascribed to TiO₂ along with the broad peak ascribed to chitosan-silica. This result indicates the effective presence of crystalline TiO₂ inside the chitosan-silica matrix.

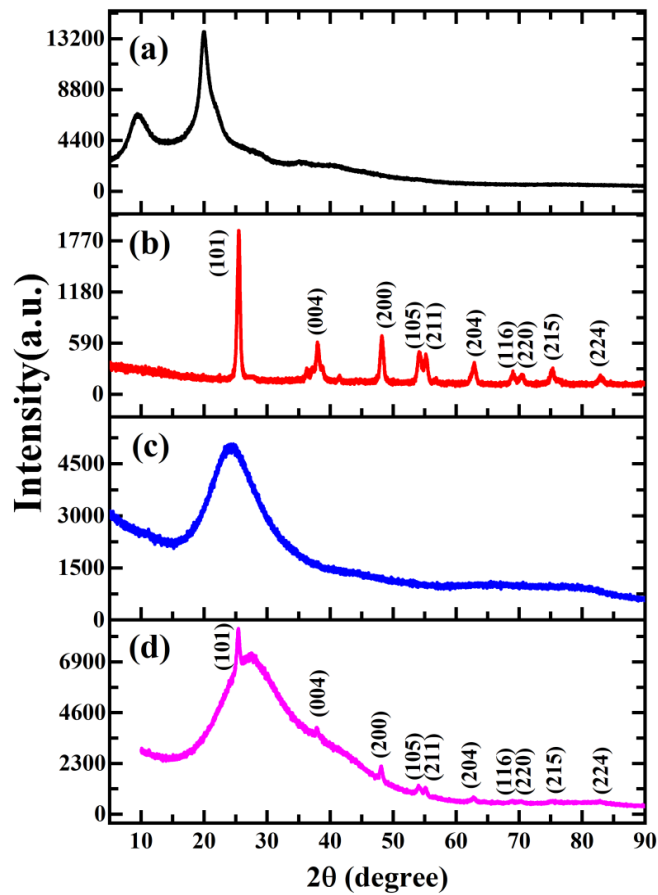

**Figure 2:** XRD plots of (a) chitosan; (b) TiO₂ NPs; (c) chitosan-silica gel; (d) chitosan-silica-TiO₂ nanocomposite gel.

### 3.2. SEM-EDX and cryo-SEM studies

The presence of $TiO_2$ NPs in chitosan-silica hybrids is also supported by SEM-EDX. Scanning electron microscopy (SEM), together with energy-dispersive X-ray spectroscopy (EDX), is a popular technique for elemental analysis. The EDX plots for chitosan-silica and chitosan-silica-$TiO_2$ nanocompositesareshown in Fig. S3 in Supporting Information. Table 1 shows the atomic weight percentages of the elements present in the chitosan-silica-$TiO_2$ hybrid obtained from SEM-EDX data. The presence of carbon, silicon, oxygen, titanium, and platinum in the composite is confirmed from the analysis. The presence of carbon can be attributed to chitosan, silicon is due to the silica, and platinum results from the coating of the platinum layer. The presence of $TiO_2$ NPs is confirmed by the presence of titanium (Ti) in the matrix. The elemental analysis table for chitosan-silica is presented in the supporting information Table ST1.

**Table 1: SEM-EDX data for elemental analysis for the chitosan-silica-$TiO_2$ hybrid**

| Element | Weight (%) | Atomic (%) |
|---|---|---|
| C | 2.41 | 3.57 |
| O | 82.91 | 92.28 |
| Si | 4.01 | 2.54 |
| Ti | 2.24 | 0.83 |
| Pt | 2.41 | 3.57 |

The surface morphologies of the chitosan-silica and chitosan-silica-$TiO_2$ nanocomposites are visualized using cryo-SEM imaging. Cryo-fractured samples, coated with a thin metallic layer of platinum, are used for this study. Fig. 3(a) displays the micro-structure of the chitosan-silica hybrid. The networks are formed by the long chains of the Si-O-Si molecules[23]. The pores of the networks are filled with chitosan. Fig. 3(b) represents the network formation by the chitosan-silica-$TiO_2$ nanocomposite. The morphological changes in the gels, particularly the increase in porosity of the network due to the incorporation of $TiO_2$, are clear from the images. An analysis with ImageJ shows that the chitosan-silica-TiO$_2$ nanocomposite has a higher average pore diameter (~ 2.37µm) than the chitosan-silica hybrid (average pore diameter ~ 1.29 µm). This feature should facilitate an improved adsorption capacity of the chitosan-silica-TiO$_2$ nanocomposite gel when compared to the chitosan-silica hybrid.

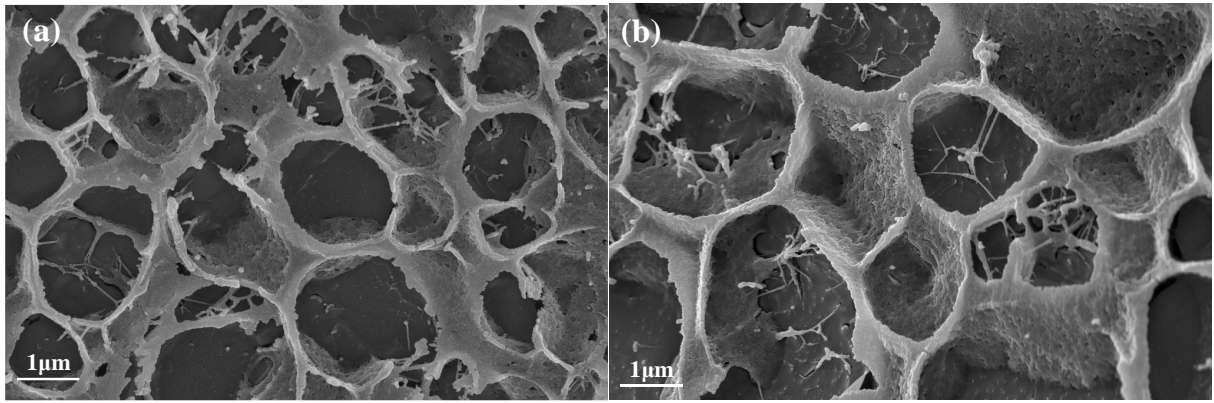

**Figure 3:** (a) A cryo-SEM image of a chitosan-silica composite and (b) chitosan-silica-TiO$_2$ nanocomposite gel.

### 3.3. Rheological Study

Rheological studies are performed to investigate the viscoelasticity of chitosan-silica hybrids upon the incorporation of TiO$_2$ NPs. Fig. 4(a) shows the mechanical responses of the chitosan-silica-TiO$_2$ nanocomposites (for varying concentrations of TiO$_2$ NPs within the range 10-100mg/20ml) with varying amplitudes of an oscillatory strain at a fixed frequency of 0.5 rad/s. The elastic or storage moduli, $G'$, and the viscous or loss moduli, $G''$, are plotted *vs.* strain amplitude $\gamma_o$ in Fig. 4(a). The maximum value of strain up to which the storage moduli remains invariant defines the upper limit of the linear viscoelastic regime (LVE) [24]. All the chitosan-silica-TiO$_2$ nanocomposites (for the varying concentration of TiO$_2$ NPs) possess higher $G'$ values than the pure chitosan-silica hybrid, indicating an improvement in mechanical rigidity of the chitosan-silica hybrid due to incorporation of TiO$_2$ NPs. Beyond the linear viscoelastic region, we note the onset of a yielding regime, wherein $G'$ decreases with increase in strain and $G''$ displays a peak above a yield strain value $\gamma_y$. The calculation of yield strain, following a protocol suggested in Ref [25], is presented in section S.B. in Supporting Information. It is noted that the values of the linear modulus $G_l'$ ($G_l'$ values at a low amplitude strain 0.5%) of all the TiO$_2$ NPs containing nanocomposites are significantly higher than that of the pure chitosan-silica composites (Fig. 4(b)). This feature is also accompanied by lower values of yield strain ($\gamma_y$) for the nanocomposites containing TiO2 NPs when compared to the chitosan-silica composite. This behavior is indicative of enhanced rigidity and the earlier onset of plastic deformations in the nanocomposites gels containing TiO$_2$ NPs. At even higher strains, $G'$ and $G''$ both decrease (Fig 4(a)), indicating that microstructures that constitute the composites are destroyed at these high strains.

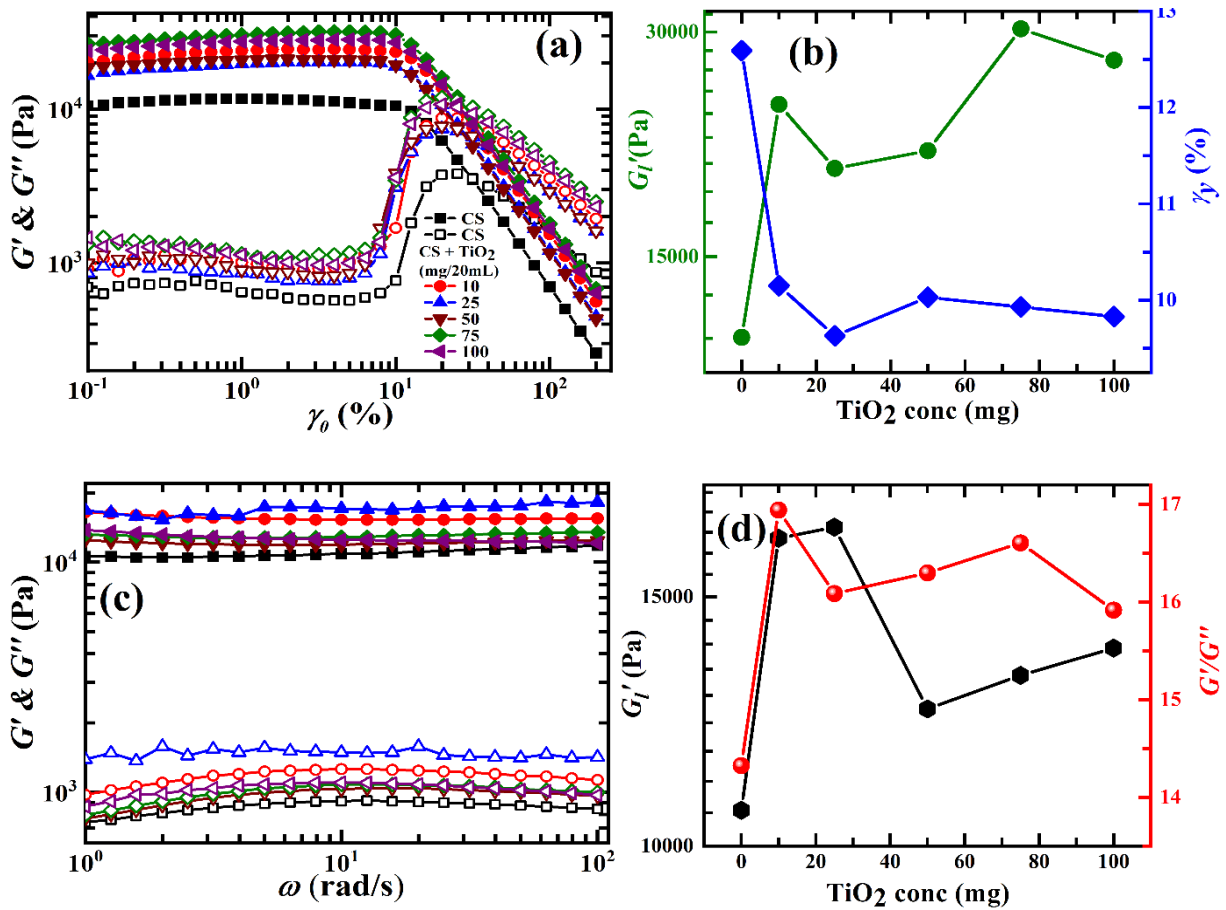

**Figure 4:** $G'$ and $G''$ are presented by solid symbols and hollow symbols respectively in Figures 4(a) and (c); (a) Strain amplitude sweep study of chitosan-silica and chitosan-silica-TiO$_2$ nanocomposites with varying concentrations of TiO$_2$ NPs (10-100 mg/20mL); (b) Low amplitude storage modulus $G_l'$ (the linear modulus, left y-axis) at varying concentrations of TiO$_2$ NPs (10-100 mg/20mL), and $\gamma_y$ at varying concentrations of TiO$_2$ NPs (right y-axis); (c) Frequency sweep study of chitosan-silica and chitosan-silica-TiO$_2$ nanocomposite with varying concentrations of TiO$_2$ NPs (10-100mg/20mL, symbols same as in (a)); (d) Low frequency storage modulus $G_l'$ at varying concentrations of TiO$_2$ NPs (10-100mg/20mL, left y-axis) and $G'/G''$ at varying concentrations of TiO$_2$ NPs (right y-axis).

All the frequency sweep studies (Fig. 4(c)) are performed at a fixed amplitude of oscillatory strain (1%), while the frequency is varied from 1-100 rad/s. The elastic moduli ($G'$) are almost independent of frequency and much higher compared to the viscous moduli ($G''$). Similar to the amplitude sweep test, the $G'$ values of chitosan-silica-TiO$_2$ nanocomposites are higher than the $G'$ value of the chitosan-silica hybrid. Fig. 4(d) shows the linear modulus $G_l'$ ($G'$ values at a low-frequency of 1rad/s) for varying concentrations of TiO$_2$ NPs. Again, there is a rapid increase in $G_l'$ upon the incorporation of TiO$_2$ though no systematic variation in $G'$ is observed with increasing TiO$_2$ NP concentration. The ratio of storage and loss modulus ($G'/G''$) is also plotted with respect to varying concentrations of TiO$_2$ (Fig. 4(d)). A significant enhancement in the sample rigidity upon the incorporation of TiO$_2$ NPs is verified.

### 3.5. Adsorption studies

The absorbance spectra of methylene blue are recorded in the concentration range 1-20mg/L at pHs 1.8-10 (Figs. S1(a),(c),(e),(g),(i),and (k) in Supporting Information). MB shows an absorbance peak at a wavelength of 664nm. Fig. 5(a) displays the dependence of absorbance peak height *vs.* MB concentrations at pH-7. Such dependences of absorbance peak heights on MB concentrations at all pHs are shown in Figs. S1(b),(d),(f),(h),(j),and (l) in Supporting Information. The linear dependence in Fig. 5(a) is indicated by a black dashed line noted at a concentration <10mg/L. The linear fit in Fig.5(a) (black dashed line) is used as a calibration curve to determine unknown MB concentration ($C_t$) after treatment with the nanocomposite, with the MB concentration values interpolated from the absorbance peak heights of treated samples. The determination of removal % is discussed in section 2.5.

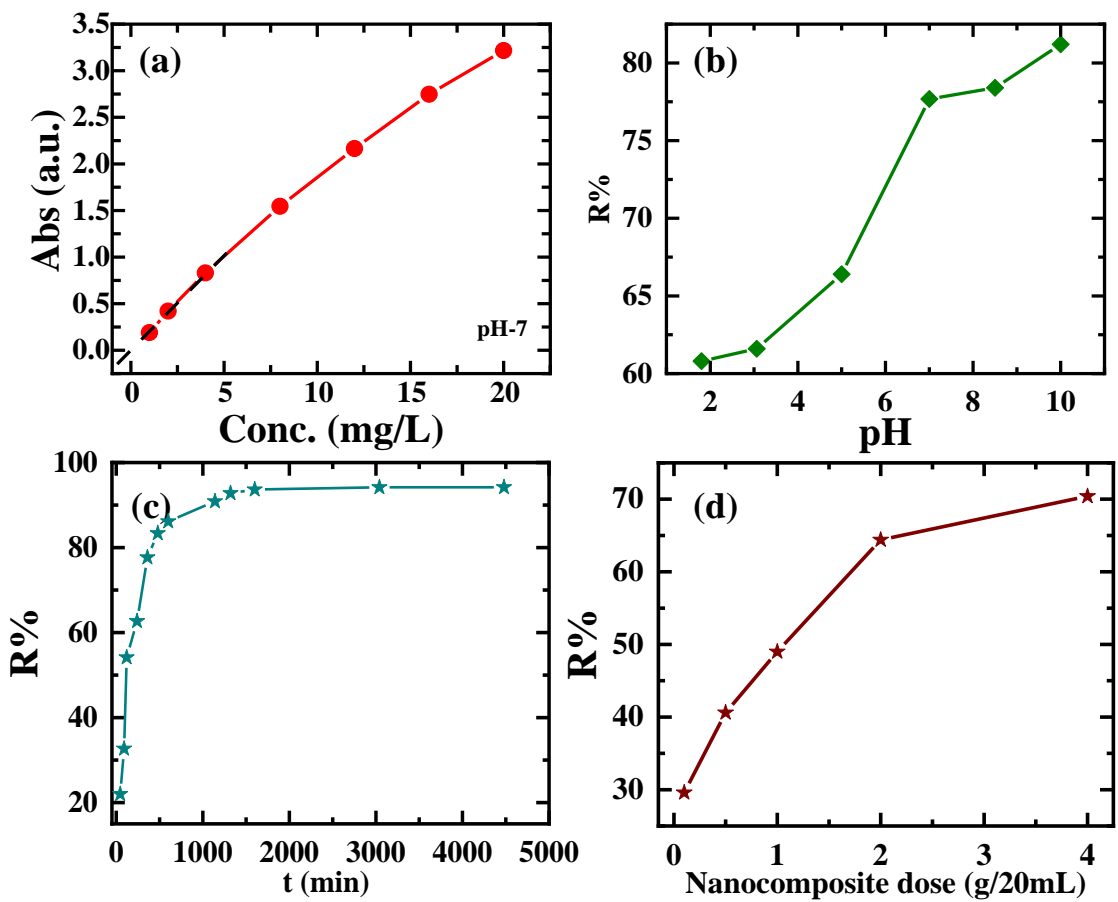

**Figure 5:** (a) Calibration plot of the absorbance values of MB at 664 nm with increasing concentration of MB (1-20mg/L) at pH-7 before treatment; (b) Effect of pH on the removal % of MB (concentration 5mg/L, contact time 400min); (c) Effect of contact time on the removal % of MB (concentration 5 mg/L, at pH-7); (d) Effect of nanocomposite dose on removal % of MB (concentration 5 mg/L, at pH-7, contact time 200 min).

The effect of the initial pH of the solution on the adsorption of MB by the chitosan-silica-TiO$_2$ nanocomposite is probed. The pH of the solutions is changed using HCl and NaOH solutions (details given in Table ST2 in Supporting Information). The concentration of TiO$_2$NPs is fixed at 50mg/20mL in the chitosan-silica solution for the rest of the adsorption studies. The adsorption study is performed under the following optimal conditions: treatment time 400min., MB dye conc. 5mg/L, chitosan-slica-TiO$_2$nanocomposite dose 1gm/20mL. Fig. S2(a) and Fig. S2(b) of Supporting Information show the absorbance spectra of MB (5mg/L) before and after the treatment with nanocomposite for a pH range of 1.8-10. The unknown concentrations ($C_t$) after the treatment are determined from the calibration plots for pHs 1.8-

10 in Figs. S1(b),(d),(f),(h),(j),(l) of Supporting Information. The removal % is calculated from Eq. 2 and plotted with respect to the varying initial pH of the solution in Fig. 5(b). The removal % increases as the initial pH is increased from 1.8 to 7 and almost saturates till pH 8.5 and again reaches almost 82% at pH 10. We believe that the increased efficiency of the nanocomposite in removing the dye at higher pH values is due to the attractive electrostatic interaction between the cationic MB dye and the anionic $TiO_2$ NPs present in the chitosan-silica-$TiO_2$ nanocomposite [22]. Since the $TiO_2$ NPs are more negatively charged at higher pH [22], the removal % is seen to increase monotonically with an increase in initial pH. It is to be noted that studies to observe the effect of contact time and the effect of nanocomposite dose on the adsorption of MB are performed at initial pH-7.

The effect of contact time on MB dye adsorption by chitosan-silica-$TiO_2$ nanocomposite is next studied at different intervals (0-4480min) at pH-7; MB dye conc.is 5mg/L; nanocomposite dose is 1gm/20mL. The absorbance spectra after treatment at various time intervals are shown in Fig.S5 in Supporting Information. The removal %, computed using Eq.2, from the acquired absorbance data, increases from 10-90% with an increase in contact time between 45-1140min. The equilibrium in adsorption is achieved at approximately around 1600min, after which the rate of adsorption reaches an equilibrium. This phenomenon is believed to arise due to the formation of a monolayer of dye molecules at the nanocomposite surface, which slows down the subsequent adsorption rate and no more adsorption sites are available for dye molecules.

The effect of chitosan-silica-$TiO_2$ nanocomposite dose on the removal % of MB dye is performed using 5mg/L MB dye solution, at pH 7 and at adsorption time of 200 min. Fig.S6 in Supporting Information displays the absorbance spectra of MB after the treatment for various nanocomposite doses. The nanocomposite dose is varied from 0.1g/20mL to 4g/20mL. The increase in the removal % with the amount of dose is presented in Fig. 5(d).

The removal %, computed using the acquired absorbance data and Eq. 2, is seen to increase from 25% to 70% as the nanocomposite dose is increased from 0.1-4g/20mL. This significant increase of removal % with increase of nanocomposite dose is attributed to the enhanced availability of surface area active adsorption sites on the nanocomposite surface at higher doses.

### 3.6. Adsorption kinetics

The evolution of the adsorption process *vs.* time is reflected by the adsorption kinetics. In the batch method, several models are used to describe the adsorption by the surface and also by the pores of the adsorbent. Among several kinetic models which are used to fit the experimental data, the most widely used ones are the pseudo-first-order (PFO) and pseudo-second-order (PSO) models and intraparticle diffusion model (IPD).

The expressions for PFO and PSO are determined by integrating the general equation [26]:

$$\frac{dq_t}{dt} = k_n(q_e - q_t)^n \qquad \text{(Eq. 3)}$$

where, $q_e$ is the adsorption capacity at equilibrium, $q_t$ is the adsorption capacity at contact time $t$ (min), and $k_n$ is the rate constant of the $n^{\text{th}}$ order model.

The pseudo-first-order model considers that the rate of occupation of adsorption sites by the adsorbate (methylene blue) on the adsorbent (chitosan-silica-$TiO_2$ nanocomposite) depends on the number of unoccupied sites. The model also assumes that adsorption happens on localized sites, and there is no interaction between adsorbed ions, which corresponds to monolayer adsorption of the adsorbed ions on the adsorbent surface. Lagergren proposed the expression of PFO model for $n = 1$ in Eq.-3 as follows [26][27]

$$\frac{dq_t}{dt} = k_1(q_e - q_t) \qquad \text{(Eq. 4)}$$

Integrating the above Eq. for the boundary conditions ($t = 0$, $q_t = 0$, and $t = t$, $q_t = q_e$) generates the following equation

$$\log(q_e - q_t) = \log q_e - \left(k_1/2.303\right) t \qquad \text{(Eq. 5)}$$

In Eq. 5, $q_e$ is the adsorption capacity at equilibrium, $q_t$ is the adsorption capacity at contact time $t$ (min) and $k_1$ (min$^{-1}$) is the kinetic rate constant of pseudo-first-order (PFO) kinetics. The linear plot of $\log(q_e - q_t)$ against time is used to determine $k_1$ (Fig. 6(a)). The fitting parameters obtained for our data are displayed in Table 2.

The pseudo-second-order (PSO) model also explains the dependence of the adsorption capacity of the adsorbent on time. While the PFO model predicts the initial stage of adsorption, the PSO model describes the adsorption behavior more accurately during the later stage of the adsorption process [27]. The PSO model assumes that the reaction rate depends on the amount of adsorbate/solute on the surface of the adsorbent. The expression for PSO was proposed by Ho and McKay and is given by considering $n = 2$ in Eq-3 as follows [26]

$$\frac{dq_t}{dt} = k_2 (q_e - q_t)^2 \qquad \text{(Eq. 6)}$$

Integrating the above Eq. for the boundary conditions ($t = 0$, $q_t = 0$, and $t = t$, $q_t = q_e$) generates the following equation

$$\left(t/q_t\right) = 1/{k_2 q_e^2} + t/q_e \qquad \text{(Eq. 7)}$$

where, $q_e$ is the adsorption capacity at equilibrium, $q_t$ is the adsorption at contact time $t$ (min) and $k_2$ (gm mg$^{-1}$ min$^{-1}$) is the kinetic rate constant of pseudo-second-order. The linear fit to $\left(t/q_t\right)$ and time data is used to determine $k_2$. The data is displayed in Fig. 6(b).

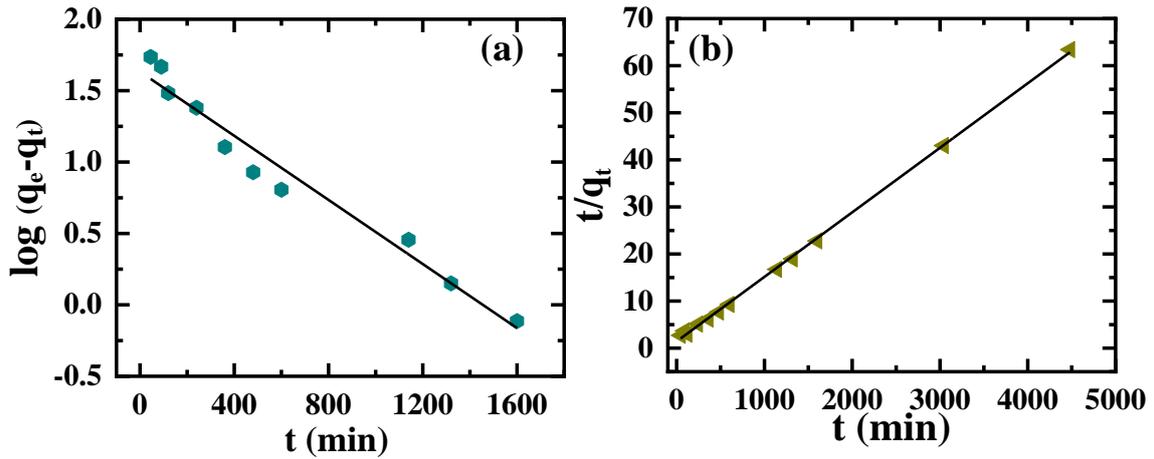

**Figure 6:** Fits to the (a) pseudo-first-order, (b) pseudo-second-order kinetic models for the adsorption of MB dye on the chitosan-silica-TiO$_2$ surface.

Fig. 6(a) and (b) are obtained from the data for the effect of contact time on adsorption plotted earlier (Fig. 5(c)). The kinetic rate constant values and the regression coefficients ($R^2$) obtained from the fits are presented in Table 2. The values for $R^2$ for the pseudo-second-order is higher than that of pseudo-first-order (PSO) implying pseudo-second-order model describe the adsorption well and adsorption is a chemisorption process.

**Table 2: Adsorption kinetics parameters from PFO and PSO models.**

| 1$^{st}$ order kinetics | | 2$^{nd}$ order kinetics | |
|---|---|---|---|
| $k_1$ (min$^{-1}$) | $R^2$ | $k_2$ (g mg$^{-1}$ min$^{-1}$) | $R^2$ |
| 0.003 | 0.96 | 1.3×10$^{-4}$ | 0.99 |

In the adsorption studies where there is a possibility of intraparticle diffusion becoming the rate-limiting step, the itraparticle diffusion model is applied [28]. This model was proposed and developed by Weber and Moris (1963) and is given by [22]

$$q_t = k_{dif} t^{1/2} + C \qquad \text{(Eq. 8)}$$

where, $q_t$ is defined as in Eq. 7 and $C$ is the intercept of the linear fit of $q_t$ vs $t^{1/2}$. The kinetic rate constant $k_{dif}$ can be determined from the slope of the plot of $q_t$ vs $t^{1/2}$. In this model, $C$ (in Eq. 8) is an arbitrary constant signifying the thickness of the boundary layer, with a larger value of $C$ corresponding to a thicker boundary layer [29]. For $C = 0$, that is, in the absence of a boundary layer, the film diffusion can be ignored and intra-particle diffusion becomes the rate controlling mechanism through the entire adsorption process. However, many studies have shown non-zero intercepts, demonstrating that the rate-limiting step involves both intra-particle and film diffusion. The data is plotted and fitted to this model in Fig. 7.

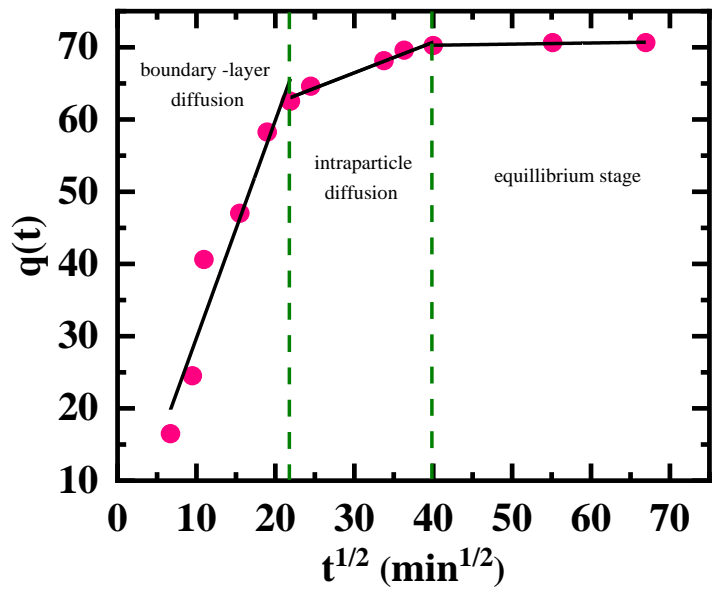

**Figure 7:** Intraparticle diffusion kinetic models for the adsorption of MB dye on the chitosan-silica-TiO$_2$ surface.

Fig. 7 shows three separate slopes, i.e., multilinearty. This plot is devided into three regions: (i) external mass transfer across the liquid boundary layer film outside the composite; (ii) the gradual adsorption stage where the intraparticle diffusion of the MB dye onto the adsorbent active sites occurs to achieve the rate-limiting step; (iii) the equilibrium stage where the adsorption sites are saturated. All the three regions of the fitted curve are linear which do not pass through the origin which suggests that adsorption is controlled by more than one mechanism. The parameters of the internal diffusion are listed in Table 3.

**Table 3: Adsorption kinetics parameters from intraparticle diffusion model.**

| Boundary-layer diffusion | | Intraparticle diffusion | | Equilibrium stage | |
|---|---|---|---|---|---|
| $k_b$ (mg g$^{-1}$ s$^{-1/2}$) | $C$ | $k_{dif}$ (mg g$^{-1}$ s$^{-1/2}$) | $C$ | $k_{eq}$ (mg g$^{-1}$ s$^{-1/2}$) | $C$ |
| 3.11 | -0.41 | 0.42 | 53.76 | 0.02 | 69.65 |

### 3.7. Adsorption isotherm

Adsorption equilibrium isotherm is fundamental in analyzing the distribution of the adsorbate molecules between the solvent and the adsorbent as the adsorption reaches an equilibrium state. The interaction mechanism of MB dye with the chitosan-silica-TiO$_2$ nanocomposite can therefore be further understood by studying the adsorption isotherm. The adsorption isotherm of the binding mechanism is studied here using the four isotherms, namely Freundlich isotherm, Dubinin–Radushkevich (D–R) isotherm, Temkin isotherm, and Langmuir isotherm. The applicability of these models are briefly discussed later.

The adsorption isotherm can be calculated based on the data for removal percentage of MB dye by chitosan-silica-TiO$_2$ nanocomposite for different initial MB dye concentrations. The absorbance spectra for different initial MB dye concentration in the range 1-16mg/L is shown in Fig. S8. The absorbance peak values (at 664nm) are plotted in Fig. 8 for 1-16mg/L MB dye concentrations.

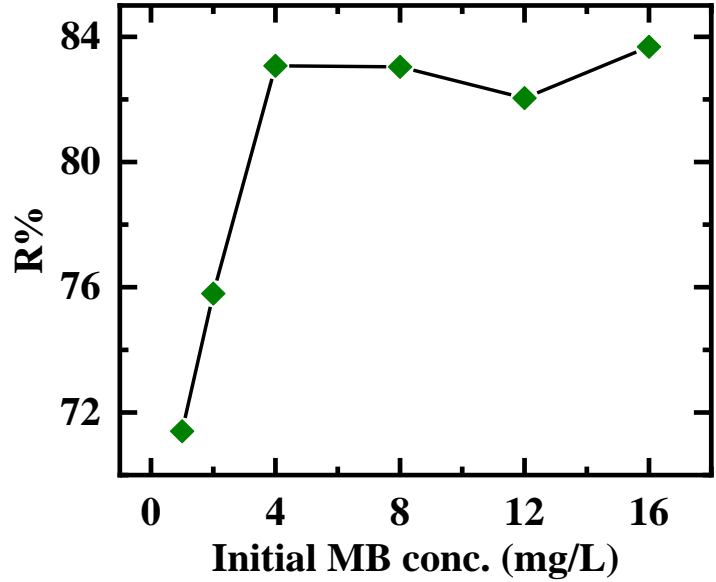

**Figure 8:** Effect of initial MB conc. on removal % when treated with chitosan-silica-$TiO_2$ nanocomposite (MB concentration 5 mg/L, at pH-7, contact time 3000 min).

The Freundlich isotherm is applied to describe an adsorption process that takes place on heterogeneous surfaces and also on active sites with different energies based on multilayer adsorption[30]. The Freundlich isotherm in linear form can be written as follows [13]

$$\log q_e = \frac{1}{n}\log C_e + \log P \qquad \text{(Eq.-9)}$$

where $q_e$ and $C_e$ are as same as in Eq-9, $P$ is the constant indicating the adsorption capacity (mg/g), and $n$ is a constant that measures the adsorption intensity. This isotherm Fig. 9(a)

The Dubinin–Radushkevich (D–R) isotherm model was developed to take into account the effect of porous structures of adsorbents [31]. The model is derived from an adsorption potential theory and the model assumes that the adsorption process is in connection with micropore volume filling in contrast to layer-by-layer adsorption on the pore walls [32]. The linear form of D-R model can be expressed as

$$log Q_e = log Q_m - \beta\varepsilon^2 \qquad \text{(Eq.10)}$$

Where, $Q_e$ is the amount of adsorbate adsorbed per unit weight of adsorbent at equilibrium (mgg$^{-1}$), $Q_m$ is the maximum adsorption capacity of adsorbent (mg/g), β is the constant related to adsorption energy, and ε is the adsorption potential (kJ$^2$mol$^{-2}$). The adsorption potential ε is defined as the following [31]

$$\varepsilon = RT ln(1 + \frac{1}{C_e}) \quad \text{(Eq. 11)}$$

Where, R is gas constant, T is the temperature in (K), $C_e$ is the equilibrium con-centration of the adsorbate. This isotherm is also shown in Fig. 9(b)

Temkin isotherm model takes into account the indirect interaction of adsorbent-adsorbate. This model assumes that the heat of adsorption of all molecules decreases linearly. Ignoring extremely low and large value of concentration, the Temkin isotherm is valid for an intermediate range of ion concentration [33]. The linear form of this model can be written as

$$q_e = \frac{RT}{b_T} ln K_T + \frac{RT}{b_T} ln C_e \quad \text{(Eq. 12)}$$

Where, $K_T$ is the equilibrium binding constant (Lmg$^{-1}$), $b_T$ is the Tempkin isotherm constant related to variation of adsorption energy (Jmol$^{-1}$). This model is shown in Fig. 9(c)

The Langmuir model assumes that the adsorption takes place at particular homogeneous sites within the adsorbent and considers that the adsorbent surface is mono-layered with uniform energy when the adsorption occurs. The Langmuir isotherm can be expressed by the linear equation [22]:

$$C_e/q_e = 1/{bq_{max}} + C_e/q_{max} \quad \text{(Eq. 13)}$$

where $b$ (L/mg) is the Langmuir model adsorption constant, $q_{max}$(mg/g) is the maximum adsorption capacity. This isotherm is shown in Fig. 9(d)

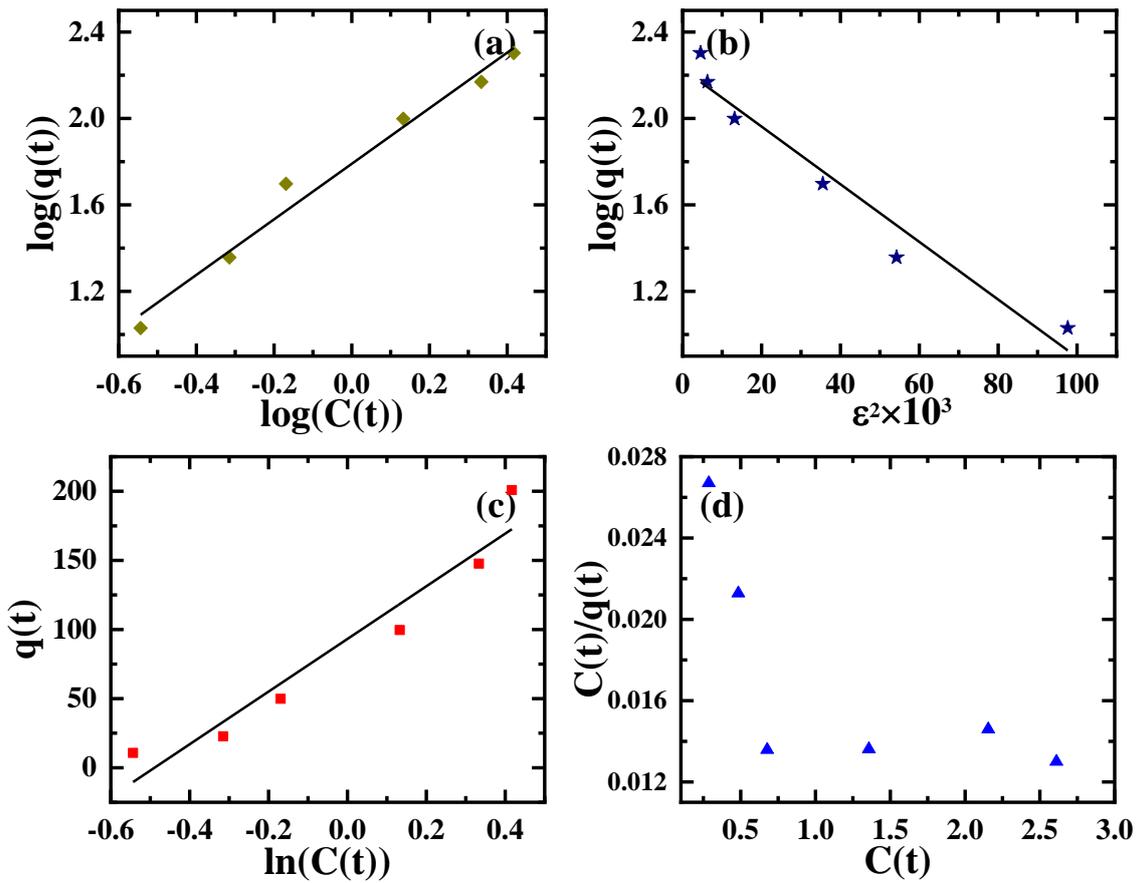

**Figure 9:** (a) Freundlich isotherm plot; (b) Dubinin–Radushkevich (D-R) isotherm plot; (c) Temkin isotherm plot; and (d) Langmuir isotherm plot.

All the isotherms are linearly fitted except Langmuir isotherm (because of its non-linearity Fig.8(d)). Freundlich isotherm model gives the best fit among the described model ($R^2 = 0.97$). Therefore, Freundlich isotherm can best describe the adsorption of MB onto chitosan-silica-TiO$_2$ nanocomposite.

**Table 3: Different isotherm parameters**

| Freundlich isotherm | | Dubinin–Radushkevich isotherm | | Temkin isotherm | |
|---|---|---|---|---|---|
| $R^2$ | 0.97 | $R^2$ | 0.95 | $R^2$ | 0.92 |
| $P$(mg/g) | 61.6 | $Q_{max}$(mg/g) | 169.43 | $K_T$(Lmg$^{-1}$) | 1.63 |
| $n$ | 0.778 | $\beta$ | $1.33 \times 10^{-5}$ | $b_T$(Jmol$^{-1}$) | 0.915 |

Finally, we note that the present nanocomposite shows promising results when compared to other chitosan composites for methylene blue adsorption. Adsorption capacities and the adsorption mechanisms for the present composite and for other reported chitosan composites for various dye adsorptions from wastewater are listed in Table 4.

**Table 4: Comparison with other reported study**

| Adsorbent | Adsorbate | Adsorption capacity | pH | Temperature | Kinetic model | Isotherm | Reference |
|---|---|---|---|---|---|---|---|
| Chitosa-silica-TiO$_2$ | Methylene Blue | 61.6 | 7.0 | 25 | Pseudo Second order | Freundlich | Present work |
| Chitosan/montmorillonite | Congo Red | 53.42 | 7.0 | 30 | Pseudo Second order | Langmuir | [34] |
| Chitosan/polyurethane | Acid Violet 48 | 30 | 7.0 | 30 | Pseudo Second order | Langmuir | [35] |
| silica gel/chitosan-g-Poly(Butyl acrylate) | Cr(VI) ions | 55.57 | 7.0 | 25 | Pseudo Second Order | Langmuir | [1] |
| N-guanidinium chitosan acetate/silica | Methylene Blue | 935 | 7.0 | - | Pseudo Second Order | Langmuir | [36] |

### 3.8 Adsorption of heavy metal ions

The chitosan-silica-TiO$_2$ composite has also been applied here for the treatment of heavy metal ions. For the present study, Cr(VI) ions are used. Cr(VI) ion solutions are prepared in the range 10-100mg/L,and the absorption spectra arerecorded (Fig. S8 in Supporting Information). The spectra display signature absorption peaks at 350 nm. The adsorption study with chitosan-silica-TiO$_2$ nanocomposite is carried out under the following conditions:the concentration of untreated Cr(VI) is taken to be 50mg/mL, pH of Cr(VI) ion solution isfixed at 7, the absorbance of the treated sample is collected after a contact time of 400min, 1g/20ml of dose of chitosan-silica-TiO$_2$ nanocomposite is used. Fig.10 represents the absorbance of the non-treated and treated Cr(VI). The decrease in the absorbance peak after treatment of Cr(VI) with the nanocomposite indicatesadsorption/removalof Cr(VI) by chitosan-silica-TiO$_2$ nanocomposite.

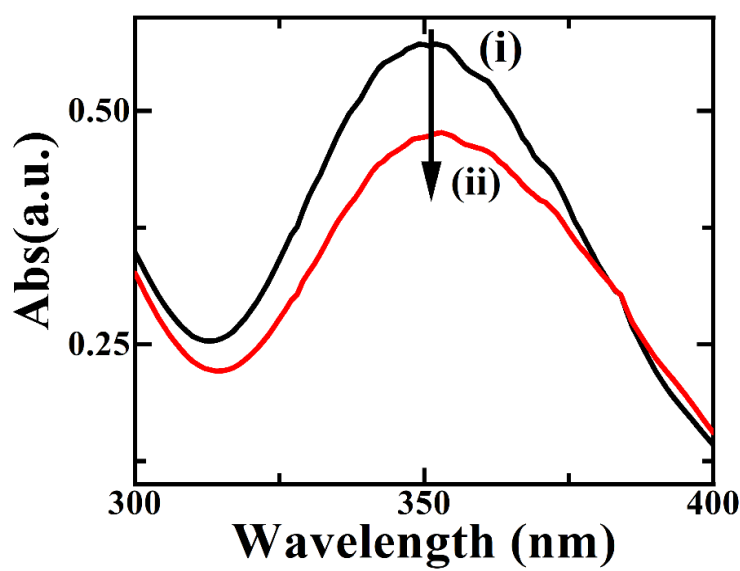

**Figure 10:** Absorbance of **(i)** non-treated Cr(VI) and **(ii)** absorbance of Cr(VI) after treatment with chitosan-silica-TiO$_2$ nanocomposite for Cr(VI) conc. 50mg/L, pH-7 and contact time 400min.

**3.8 Conclusion**

In this work, methylene blue (MB) dye is successfully adsorbed by chitosan-silica-TiO$_2$ nanocomposites, and the removal % and the maximum adsorption capacity are computed. The nanocomposite is synthesized by a sol-gel polymerization method. The formation of the nanocomposite is supported by X-ray diffraction, cryo-SEM-EDX analysis, and its improved mechanical properties are investigated using rheological measurements. Furthermore, the adsorption of MB from its aqueous solution is studied systematically. The effects of pH, contact time, and chitosan-silica-$TiO_2$ nanocomposite dose are reported. The adsorption of MB on chitosan-silica-$TiO_2$ nanocomposite is best fitted by the pseudo-second-order (PSO) kinetic model and is well-described by the Freundlich isotherm model. Thus, this work establishes chitosan-silica-$TiO_2$ as a new candidate material exhibiting an excellent dye removal capacity that can be considered as a good adsorbent material for the removal of cationic MB in water purification. We have also demonstrated the versatility of the nanocomposite gel by demonstrating its ability to remove Cr(VI) ions. To conclude, the present biocompatible chitosan-silica-$TiO_2$ nanocomposite serves as a promising, versatile adsorbent and further studies need to be performed to evaluate its potential uses in wastewater treatment.

**Acknowledgement**

We thank K M Yatheendran for his help with cryo-SEM imaging and SEM-EDX analysis and K N Vasudha for her help with the UV-visible measurements. We acknowledge Raman Research Institute and DST SERB EMR/2016/0006757 for funding this research.

# Supporting Information

**Synthesis and characterization of new chitosan-based nanocomposite gel and its application towards dye removal**

*Tamal Sarkar\**

Raman Research Institute, Bangalore, India

*Email: tamalsarkar09@gmail.com

**S.A. Determination of Removal %**

The general scheme that was used to determine removal % (Eq. 2 in the main manuscript) is discussed in this section for a dye solution at pH-7. A similar protocol is used for dye solutions at other pHs as well. All the raw absorption data in a pH range of 1.8 - 10 are provided in Fig. S1.

$$\text{Removal \%} = ((C_0 - C_t)/C_0) \times 100$$

(i) Absorbance of methelyne blue (MB) solution is recorded at pH-7 for the concentration range 1-20 mg/L (Figure S1(g) of Supporting Information). Before treatment with the nanocomposite, the absorbance peak points (at 664 nm) in the absorption spectrum for each MB concentration are plotted with respect to increasing concentration (Figure S1(h) of Supporting Information). This MB sample is called an untreated sample. This calibration plot serves the purpose of determining any unknown MB concentration if the absorbance value is given.

(ii) Next, MB solution of concentration 5mg/L is taken at pH-7 (volume 20 ml). 1g dose of chitosan-silica-TiO$_2$ nanocomposite is added to the MB solution. The system (nanocomposite dose in the MB solution) is left undisturbed for adsorption to occur. This MB sample is called a treated sample. After 400 min., the absorbance of treated MB is recorded (curves for treated and untreated MB solution at pH-7 in Figs. S2(a) and S2(b) of Supporting Information).

(iii) Once the peak absorbance of the treated MB solution is obtained (peak absorbance value at 664 nm), the concentration ($C_t$ for $t = 400$ min) of the treated MB after 400 min of adsorption is determined from the calibration plot of MB dye solution at pH-7 (Fig. S1(h) of Supporting Information).

For a representative MB concentration of $C_0 = 5$ mg/L,

Absorbance for treated MB sample at pH 7 from Fig. S2(b) at 664 nm ~0.22.

From Fig.S1(h) of Supporting Information, the unknown concentration of treated MB sample $C_{t=400\ min}$ =1.26 mg/L.

Removal % = $\left(\frac{(C_0-C_t)}{C_0}\right) \times 100$ = ((5-1.26)/5)×100 = 74.8 %

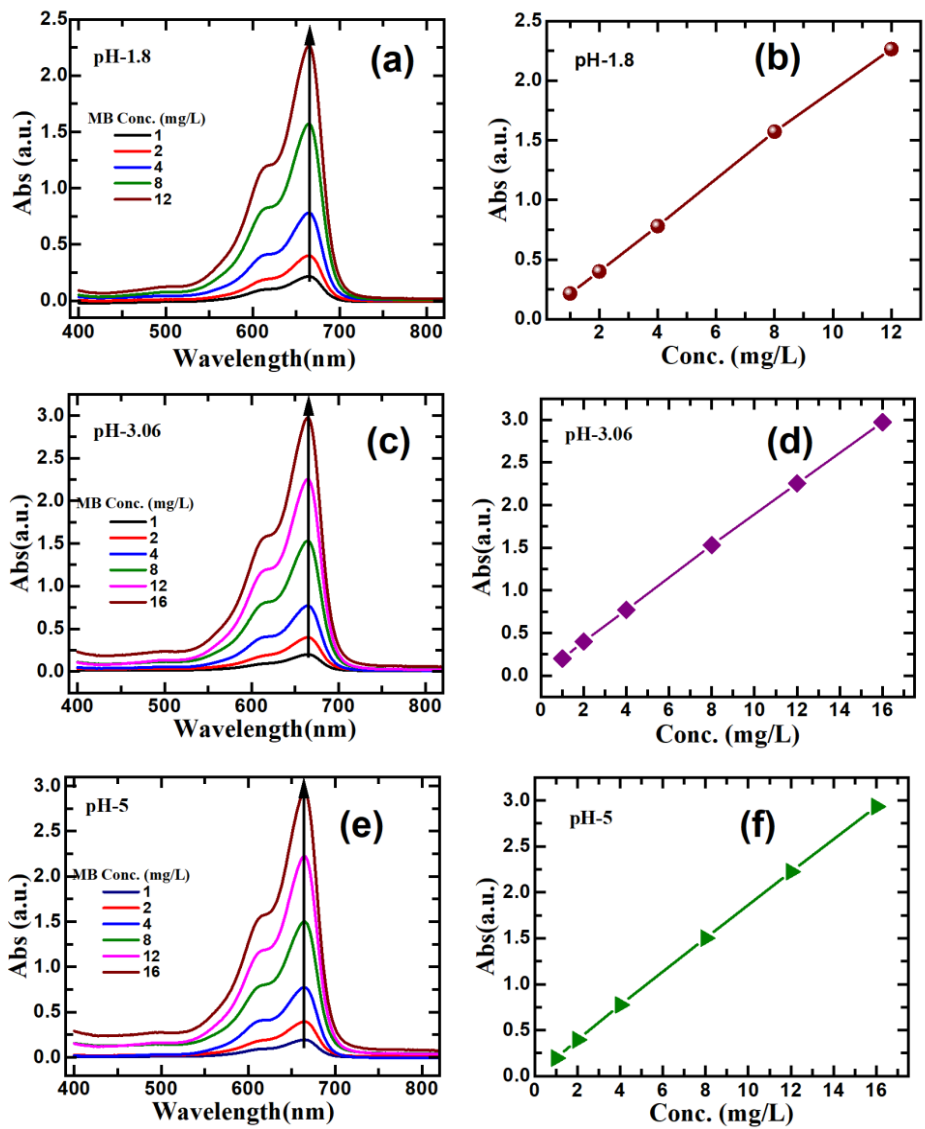

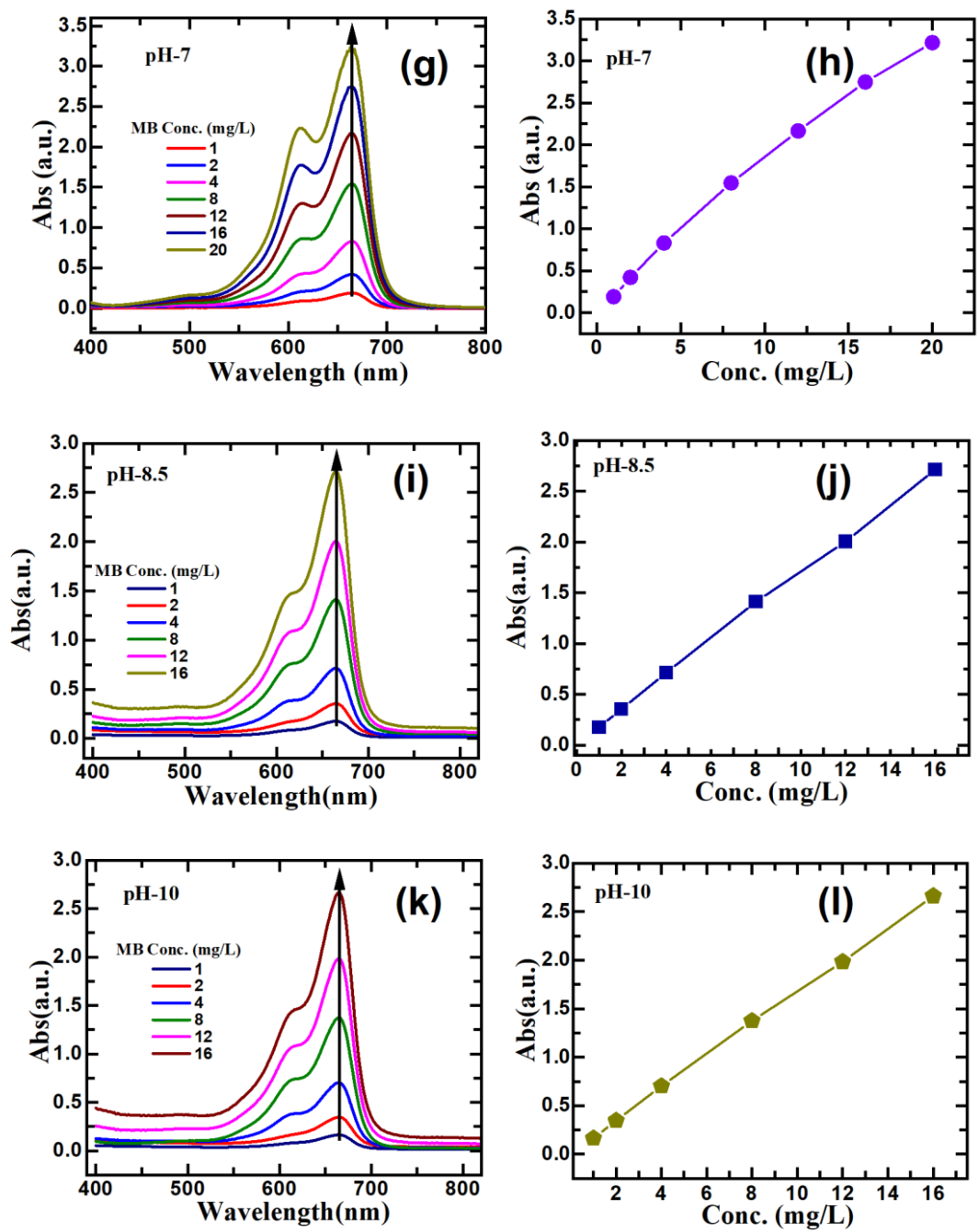

**Figure S1: (a), (c), (e), (g), (i)** and **(k)** Absorbance spectra of methylene blue dye in the concentration range 1-20 mg/L for pHs 1.8 - 10; **(b), (d), (f), (h), (j),** and **(l)** Dependence of absorbance peak value on MB concentration for pH-1.8-10

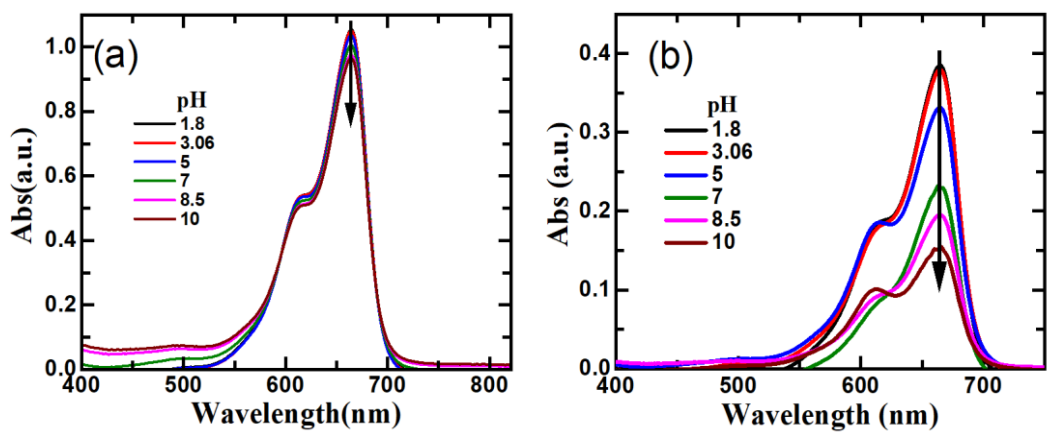

**Figure S2: (a)** Absorbance spectra of non-treated MB samples of concentration 5 mg/L at different pHs 1.8-10; **(b)** Absorbance spectra of treated MB samples at 400 min. at different pHs 1.8-10.

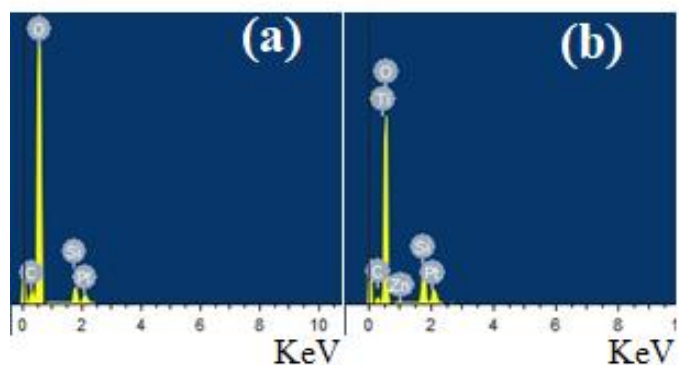

**Figure S3:** SEM-EDX plots for (a) chitosan-silica hybrid and (b) chitosan-silica-TiO$_2$ nanocomposite

**Table ST1:** Table for SEM-EDX data for elemental analysis for a chitosan-silica hybrid

| Element | Weight % | Atomic % |
| --- | --- | --- |
| C | 8.35 | 11.47 |
| O | 83.84 | 86.49 |
| Si | 2.74 | 1.61 |
| Pt | 5.07 | 0.43 |

## S.B. Determination of yield strain ($\gamma_y$)

For the determination of yield strain $\gamma_y$, a method described by Laurati *et al.* (Ref [27] in the manuscript and [1] at the end of Supporting Information) is followed. According to this method, first, the elastic stress $\sigma_{el} = G'\gamma_0$ is plotted (Figure S4 in Supporting Information) with respect to strain $\gamma_0$. The elastic stress *vs.* strain data is next fitted to $\sigma_{el} = G'\gamma_0$ in the low $\gamma_0$ range. The value of $\gamma_y$ is defined as the value of $\gamma$ at which the measured value of $\sigma_{el}$ starts to deviate by more than 3% from the Hooke's law value (linear plot of $\sigma_{el} vs. \gamma$). In case of chitosan-silica-TiO$_2$ nanocomposite (with 50 mg TiO$_2$) shown in Figure S4, the yield strain value is calculated to be $\sigma_{el} = 10.03\%$.

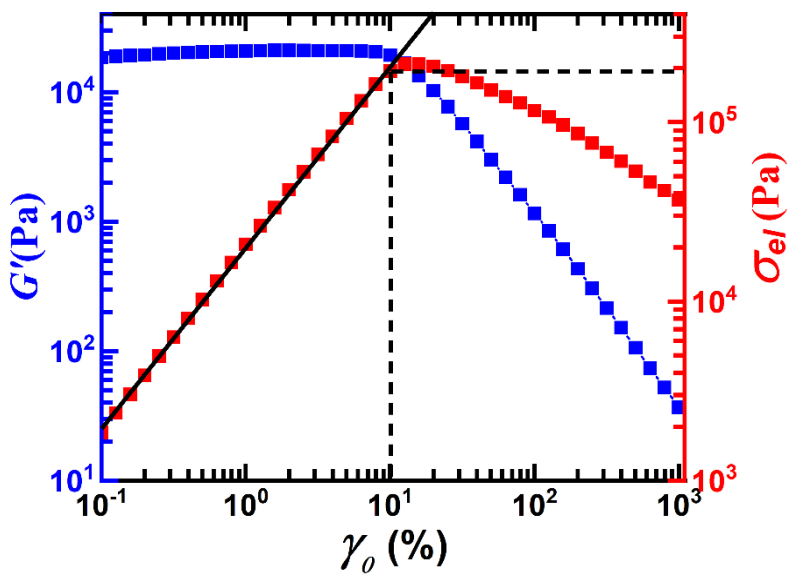

**Figure S4:** Variation of storage modulus ($G'$) and elastic stress ($\sigma_{el}$) with strain amplitude $\gamma_0$ at an angular frequency of 0.5 rad/s for a chitosan-silica-TiO$_2$ nanocomposite (with TiO$_2$ NPs concentration 50 mg/20ml). The estimation of the yield strain $\gamma_y$ is also displayed.

**Table ST2:** Procedure to change the pH for MB solution of volume 20ml (intial pH-6.65).

| pH | Volume of 0.1M HCL added | Volume of 0.001M NaOH sdded |
|---|---|---|
| 1.8 | 70µL | -- |
| 3.06 | 25µL | -- |
| 5 | 2µL | -- |
| 7 | -- | 15µL |
| 8.5 | -- | 45µL |
| 10 | -- | 60µL |

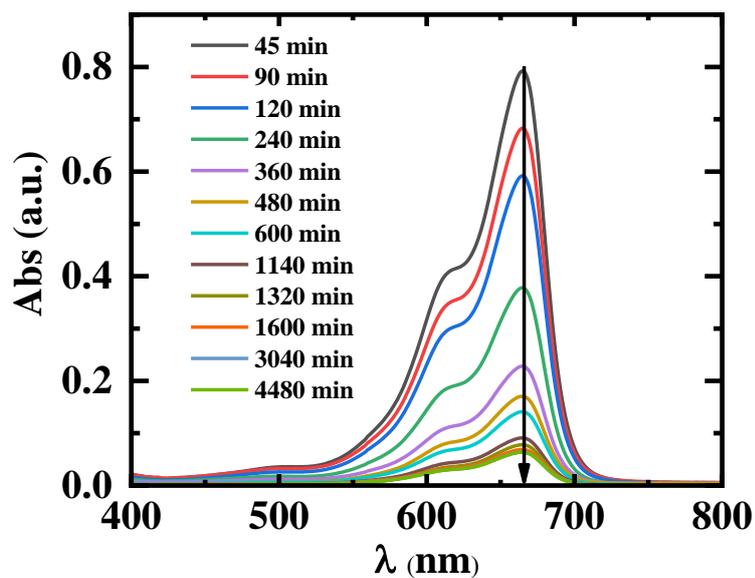

**Figure S5:** Absorbance spectra of treated 5 mg/L MB samples at various intervals (45-4480min) at pH 7.

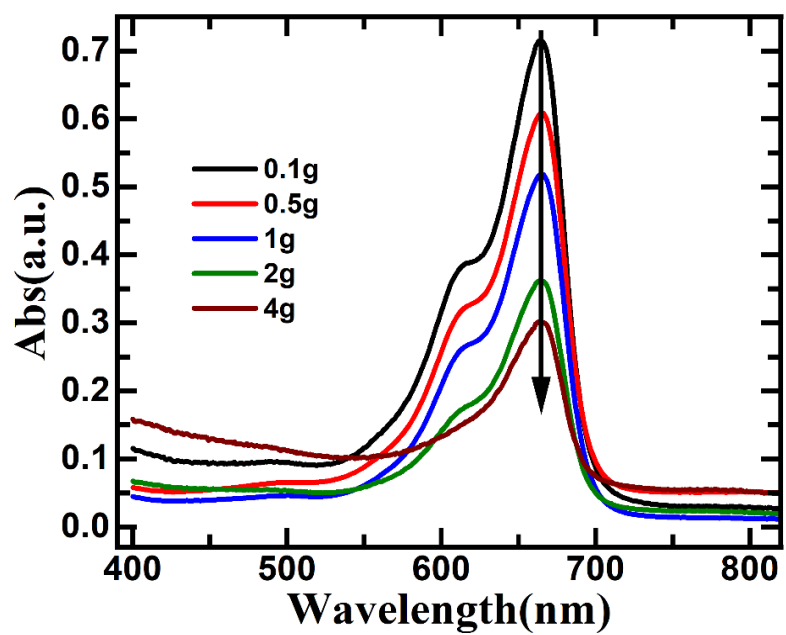

**Figure S6:** Absorbance spectra of treated MB samples for various nanocomposite doses (0.1-4 g/20mL) at pH 7.

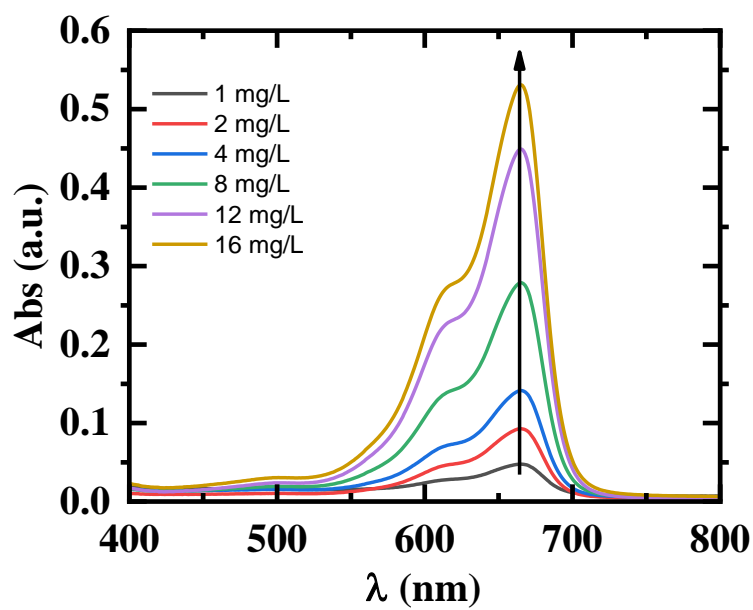

**Figure S7:** Absorbance spectra of treated MB samples for various initial MB concentrations (1-16 mg/L) at pH 7.

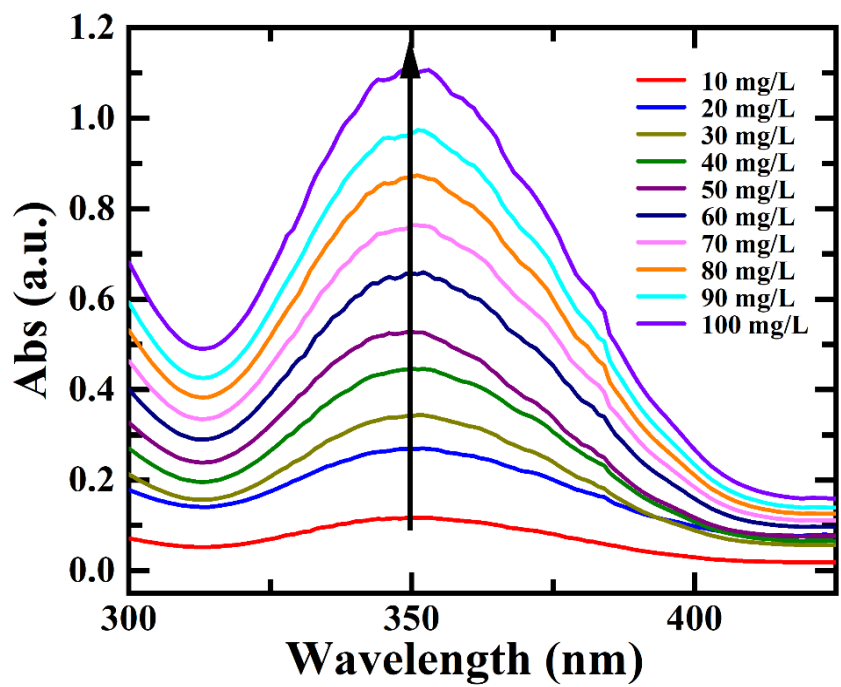

**Figure S8:** Absorbance of aqueous solutions Cr(VI) ions in the range 10-100 mg/L at pH-7.